\documentstyle[aps,prb,eqsecnum,epsf,floats]{revtex}
\begin{document}
\def\SNG{{\em Physical Review Style and Notation Guide}}
\def\LUG {{\em \LaTeX{} User's Guide \& Reference Manual}}
\def\btt#1{{\tt$\backslash$\string#1}}%
\def\REVTeX{REV\TeX}
\def\AmS{{\protect\the\textfont2
        A\kern-.1667em\lower.5ex\hbox{M}\kern-.125emS}}
\def\AmSLaTeX{\AmS-\LaTeX}
\def\BibTeX{\rm B{\sc ib}\TeX}
\twocolumn[\hsize\textwidth\columnwidth\hsize\csname@twocolumnfalse%
\endcsname

\title{Quantum critical behavior of itinerant ferromagnets}
\author{D.Belitz}
\address{Department of Physics and Materials Science Institute\\
University of Oregon,\\
Eugene, OR 97403}
\author{T.R.Kirkpatrick}
\address{Institute for Physical Science and Technology, and Department of Physics\\
University of Maryland,\\ 
College Park, MD 20742}

\date{\today}
\maketitle

\begin{abstract}
The quantum ferromagnetic transition of itinerant electrons is considered.
We give a pedagogical review of recent results which show that zero-temperature 
soft modes that are commonly neglected,
invalidate the standard Landau-Ginzburg-Wilson description of this
transition. If these modes are taken into account, then the resulting
order parameter field theory is nonlocal in space and time. Nevertheless,
for both disordered and clean systems the critical
behavior has been exactly determined for spatial dimensions $d>2$ and $d>1$,
respectively. The critical exponents characterizing the 
paramagnetic-to-ferromagnetic transition are dimensionality dependent,
and substantially different from both mean-field critical exponents, and from
the classical Heisenberg exponents that characterize the transition at 
finite temperatures. Our results should be easily observable, particularly
those for the disordered case, and experiments to check our
predictions are proposed.
\end{abstract}
\pacs{PACS numbers: }
]

\section{Introduction}
\label{sec:I}

Phase transitions that occur in a quantum mechanical system at zero 
temperature ($T=0$) as a function of some
non-thermal control parameter are called quantum phase transitions. In
contrast to their finite-temperature counterparts, which are often
referred to as thermal or classical phase transitions, the 
critical fluctuations
one has to deal with at zero temperature are quantum fluctuations
rather than thermal ones, and the need for a quantum mechanical treatment
of the relevant statistical mechanics makes the theoretical description
of quantum phase transitions somewhat different from that of classical
ones. However, as Hertz has shown in a seminal paper,\cite{Hertz} 
the basic theoretical concepts that have led to successfully 
describe and understand thermal transitions work in the quantum case as well.

Experimentally, the zero-temperature behavior of any material can of course
not be studied directly, and furthermore the
most obvious control parameter that
drives a system through a quantum transition is often some microscopic
coupling strength that is hard to change experimentally. As a result, the
dimensionless distance from the critical point, $t$, which for classical
transitions with a transition temperature $T_c$ is given by $t=T/T_c - 1$ and
is easy to tune with high accuracy, is much harder to control in the
quantum case. However, $t$
is usually dependent on some quantity that can be
experimentally controlled, like e.g. the composition of the material.
Also, the zero temperature critical behavior manifests itself already at
low but finite temperatures. Indeed, in a system with a very low thermal
transition temperature all but the final asymptotic behavior in the
critical region is dominated by quantum effects. The study of quantum
phase transitions is therefore far from being of theoretical interest only.

Perhaps the most obvious example of a quantum phase transition is the
paramagnet-to-ferromagnet transition of itinerant electrons at
$T=0$ as a function of the exchange interaction between the electronic
spins. Early theoretical work\cite{Hertz} on this transition suggested that the
critical behavior in the physical dimensions $d=2$ and $d=3$ was not
dominated by fluctuations, and mean-field like, as is the thermal
ferromagnetic transition in dimensions $d>4$. The reason for this is
a fundamental feature of quantum statistical mechanics, namely the fact
that statics and dynamics are coupled. As a result, a quantum mechanical
system in $d$ dimensions is very similar so the corresponding classical
system in $d+z$ dimensions, where the so-called dynamical critical exponent
$z$ can be thought of as an extra dimensionality that is provided to the
system by time or temperature. The $d+z$-dimensional space relevant for
the statistical mechanics of the quantum system bears some resemblance to
$d+1$-dimensional Minkowski space, but $z$ does {\em not} need to be equal to 
$1$ in nonrelativistic systems. For clean and disordered itinerant quantum
ferromagnets, one finds $z=3$ and $z=4$, respectively, in mean-field theory. 
This appears to
reduce the upper critical dimension $d_c^+$, above which fluctuations 
are unimportant and simple mean-field theory yields the correct critical
behavior, from $d_c^+ = 4$ in the classical case to $d_c^+ = 1$ and
$d_c^+ = 0$, respectively, in the clean and disordered quantum cases.
If this were true, then this quantum phase transition would be rather 
uninteresting from a critical phenomena point of view. 

It has been known for some time that, for the case of disordered
systems, this conclusion cannot be correct.\cite{Millis} It is known that
in any system with quenched disorder that undergoes a phase transition, the 
critical exponent $\nu$ that describes the divergence of the correlation
length, $\xi \sim t^{-\nu}$ for $t\rightarrow 0$, must satisfy the inequality 
$\nu\geq 2/d$.\cite{Harris}
However, mean-field theory yields $\nu = 1/2$, which is incompatible with
this inequality for $d<4$. Technically, this implies that the disorder
must be a relevant perturbation with respect to the mean-field fixed point.
The mean-field fixed point must therefore be unstable, and the phase
transition must be governed by some other fixed point that has a
correlation length exponent $\nu\geq 2/d$. 

Recently such a non-mean field like fixed point has been
discovered, and the critical behavior has been determined exactly for
all dimensions $d>2$.\cite{fm} It was found that both the value $d_c^+ = 0$
for the upper critical dimension, and the prediction of mean-field critical
behavior for $d>d_c^+$ were incorrect. Instead, $d_c^+ = 2$, and while both the
quantum fluctuations and the disorder fluctuations are irrelevant with
respect to the new fixed point for all $d > d_c^+$, there are two
other ``upper critical dimensionalities'', $d_c^{++} = 4$ and
$d_c^{+++} = 6$. The critical behavior for $d_c^+ < d < d_c^{+++}$
is governed by a non-standard Gaussian 
fixed point with non-mean field like exponents, and only for $d > d_c^{+++}$ 
does one obtain mean-field exponents. In addition, the clarification of the
physics behind this surprising behavior has led to the conclusion that
very similar effects occur in clean systems.\cite{clean} In that case,
$d_c^+ = 1$ in agreement with the early result, but again the critical
behavior is nontrivial in a range of dimensionalities 
$d_c^+ < d \leq d_c^{++} = 3$, and only for $d > d_c^{++}$ does one obtain
mean-field critical behavior. In addition, we have found that Hertz's
$1-\epsilon$ expansion for the clean case is invalid. This explains an
inconsistency between this expansion and an exact exponent relation that
was noted earlier by Sachdev.\cite{Sachdev} 
In order to keep our discussion focused, in
what follows we will restrict ourselves to the disordered case, where the
effects are more pronounced, and will only quote results for the clean
case where appropriate.

The basic physical reason behind the complicated behavior above the upper
critical dimensionality $d_c^+$, i.e. in a regime in parameter space 
where the critical behavior is not dominated by fluctuations, is simple.
According to our general understanding of continuous phase transitions
or critical points, in order to understand the critical singularities at
any such transition, one must identify all of the slow or soft modes near
the critical point, and one must make sure that all of these soft modes are
properly included in the effective theory for the phase transition. This
is obvious, since critical phenomena are effects
that occur on very large length
and time scales, and hence soft modes, whose excitation energies vanish in
the limit of long wavelengths and small frequencies, 
will in general influence the critical behavior.
In previous work on the ferromagnetic transition it was implicitly assumed
that the only relevant soft modes are the fluctuations of the order
parameter, i.e. the magnetization. For finite temperatures this is correct.
However, at $T=0$ there are additional soft modes in a disordered electron
system, namely diffusive particle-hole excitations that are distinct from
the spin density excitations that determine the magnetization. In many-body
perturbation theory these modes manifest themselves as products of retarded
and advanced Green's functions, and in field theory they can be interpreted as
the Goldstone modes that result from the spontaneous breaking of the
symmetry between retarded and advanced correlation functions, or between
positive and negative imaginary frequencies.
In a different context, namely the
transport theory for disordered electron systems, these diffusive
excitations are sometimes referred to as `diffusons' and `Cooperons',
respectively, and they are responsible for what is known as 
`weak localization effects' in disordered electron systems.\cite{WeakLoc}
For our purposes,
their most important feature is their spatial long-range nature in the
zero frequency limit. This long-range nature follows immediately from the
diffusion equation
\begin{mathletters}
\label{eqs:1.1}
\begin{equation}
\left(\partial_t - D\,\partial_{\bf x}^2\right)\,f({\bf x},t) = 0\quad,
\label{eq:1.1a}
\end{equation}
for some diffusive quantity $f$, with $D$ the diffusion constant. Solving
this equation by means of a Fourier-Laplace transform to wavevectors ${\bf q}$
and complex frequencies $z$, one obtains in the
limit of zero frequency,
\begin{equation}
f({\bf q},z=0) = {1\over D{\bf q}^2}\,f({\bf q},t=0)\quad.
\label{eq:1.1b}
\end{equation}
\end{mathletters}%
Long-range static correlations are thus an immediate consequence of the
diffusive nature of the density dynamics in disordered systems. 

The fact
that we are concerned with the zero frequency or long-time limit is due
to the order parameter, i.e. the magnetization, being a conserved quantity.
Since the only way to locally change the order parameter density is to
transport this conserved quantity from one region in space to another,
in order to develop long-range order over arbitrarily large
distances the systems needs an infinitely long time. This in turn means
that criticality can be reached only if the frequency is taken to zero
before the wavenumber. This feature would be lost if there were some type
of spin-flip scattering mechanism present, and our results hold only in the
absence of such processes. For the same reason, they do not apply to
quantum antiferromagnets, which show a quite different behavior.\cite{afm} 

It is important that the long-range static correlations mentioned above are 
distinct from the order parameter fluctuations. For instance, the latter
are soft only at the critical point and in the ordered phase, while the
former are soft even in the paramagnetic phase, and they do not change
their nature at the critical point. However, since they couple to the
conserved order paramter, they influence the critical behavior. If one
integrates out these diffusive modes in order to obtain an effective
theory or Landau-Ginzburg-Wilson (LGW) functional in terms of the order
parameter only, then their long-range 
nature leads to infrared singular integrals,
which in turn results in singular vertices in the LGW funcional, or 
diverging coupling constants for couplings between the order parameter
fluctuations. The usual LGW philosophy of deriving an effective local
field theory entirely in terms of the order parameter field therefore does
not lead to a well behaved field theory in this case. 
The situation is analogous to a well known phenomenon
in high energy physics: Suppose some interaction between, say, fermions,
is mediated by the exchange of some other particles, e.g. gauge bosons of
mass $M$. If the bosons are integrated out, then the resulting theory will
be nonrenormalizable, i.e. it will be ill-behaved 
on momentum scales larger than the
mass $M$. The nonrenormalizable theory corresponds to the order parameter
LGW theory, except that in statistical mechanics one runs into infrared
problems rather ultraviolet ones. Nevertheless, it turns out that in
our case the critical behavior can still be determined 
exactly even after having
integrated out the additional soft modes. The point is that the diffusive
modes lead to an effective long-range interaction between the order parameter
fluctuations that falls off in real space like $r^{2-2d}$. It is known that
in general long-range interactions suppress fluctuation effects.\cite{FMN} 
In our case
they are strong enough to not only suppress quantum fluctuations, but also
any remaining disorder fluctuations. The critical behavior is thus neither
dominated by quantum fluctuations (since we work above the upper critical
dimension $d_c^+$), nor by the disorder fluctuations, but rather is given
by a simple, though non-standard (because of the long-range interactions)
Gaussian theory. The resulting Gaussian fixed point allows for a correlation
length exponent that satisfies $\nu\geq 2/d$ as required, and the exponents
are dimensionality dependent for all $d<6$. In $d=3$ they are substantially
different from either the mean-field exponents, or from those for a classical
Heisenberg ferromagnet. This has striking observables consequences, as we
will discuss.

The outline of this paper is as follows. In Sec. \ref{sec:II} we first
discuss some general aspects of itinerant ferromagnets, and then we give
our results for the critical exponents and for the equation of state near
the critical point. Since the purpose of this paper is to give an exposition
and discussion of these results that is as nontechnical as possible,
they will be presented without any derivations. In Sec.\ \ref{sec:III} 
we discuss these results as well
as several possible experiments that could be performed to test our
predicitions. Finally, in Sec.\ \ref{sec:IV} we sketch the derivation of our
theoretical results.

\section{Results}
\label{sec:II}

\begin{figure}
\epsfxsize=8.25cm
\epsfysize=6.6cm
\epsffile{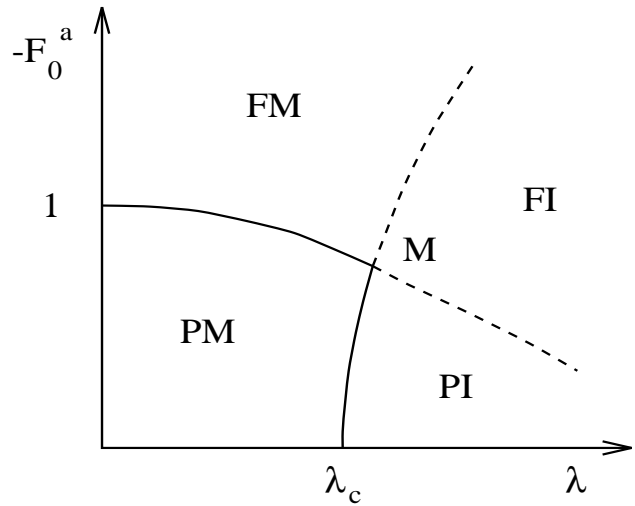}
\vskip 0.5cm
\caption{Schematic phase diagram for a $3$-$d$ disordered itinerant electron
system in the plane spanned by the Landau parameter $F_0^a$ and the
disorder $\lambda$ at $T=0$. See the text for further explanations.}
\label{fig:1}
\end{figure}

In order to put the phase transition we are going to consider in perspective,
let us first discuss the qualitative phase diagram that one expects for a
disordered itinerant electron system in $d=3$. Let $F_0^a<0$ be the
Fermi liquid parameter that characterizes the strength of the sytem's
tendency towards ferromagnetism: For $\vert F_0^a\vert < 1$ the system is
paramagnetic with a spin susceptibility $\chi_s \sim 1/(1+F_0^a)$, while
for $\vert F_0^a\vert > 1$ the clean Fermi liquid has a ferromagnetic ground
state. In Fig.\ \ref{fig:1} we show the qualitative phase diagram one
expects for a disordered system at $T=0$ in the $F_0^a$-$\lambda$ plane,
where $\lambda$ is some dimensionless measure of the disorder. For $\lambda=0$,
we have the transition from a paramagnetic metal (PM) to a ferromagnetic
metal (FM) at $F_0^a = -1$. At small but nonzero $\lambda$ this transition
will occur at somewhat smaller values of $\vert F_0^a\vert$, since the
disorder effectively increases the spin triplet electron-electron interaction
amplitude, and hence $\vert F_0^a\vert$. This accounts for the downward
curvature of the PM-FM transition line. At $\vert F_0^a\vert = 0$, a 
metal-insulator transition of Anderson type is known to occur at a critical
disorder value $\lambda_c$.\cite{LeeRama}
At nonzero $\vert F_0^a\vert$ such a transition
from a paramagnetic metal to a paramagnetic insulator (PI) still occurs,
albeit it now is what is called an Anderson-Mott transition that takes place
at a somewhat larger value of the disorder.\cite{R}
The two transition lines will
meet at a multicritical point M, and for
large values of $\lambda$ and $\vert F_0^a\vert$ one expects a ferromagnetic
insulator (FI). The transitions from the FM and PI phases, respectively,
to the FI phase have not been studied theoretically, which is why we denote
them by dashed lines in the figure.
We will be mostly interested in the phase transition that occurs across 
the PM-FM transition line at finite disorder, but far away from the 
metal-insulator transition. However, in Sec.\ \ref{sec:III} below 
we will come back to the remaining regions in this phase diagram.

\begin{figure}
\epsfxsize=8.25cm
\epsfysize=6.6cm
\epsffile{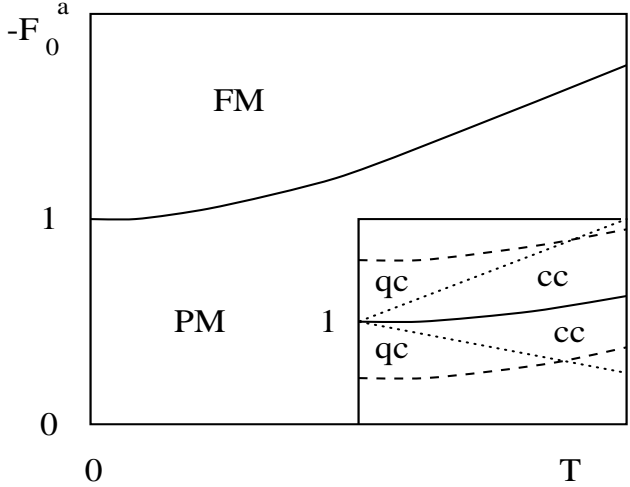}
\vskip 0.5cm
\caption{Schematic phase diagram for a disordered itinerant electron
system in the plane spanned by the Landau parameter $F_0^a$ and the
temperature $T$. The inset shows the boundary of the critical region
(dashed line) and the crossover line (dotted line) that separates
classical critical behavior (cc) from quantum critical behavior (qc).}
\label{fig:2}
\end{figure}

In Fig.\ \ref{fig:2} we show the same phase diagram in the 
$F_0^a$-$T$ plane for some value of the disorder $0 < \lambda << \lambda_c$.
With increasing temperature $T$, the critical value of $\vert F_0^a\vert$
increases, since in order to achieve long-range order, a larger 
$\vert F_0^a\vert$ is needed to compensate for the disordering effect of
the thermal fluctuations. The inset shows schematically the boundary of the
critical region (dashed line) and the crossover line between classical and
quantum critical behavior (dotted line). At any nonzero $T$, the asymptotic
critical behavior is that of a classical Heisenberg magnet, but at sufficiently
low $T$ there is a sizeable region where quantum critical behavior can be
observed.

Our theoretical results for the zero temperature paramagnet-to-ferromagnet
transition can be summarized as follows. Let $t$ be the dimensionless
distance from the line separating the regions PM and FM in Fig.\ \ref{fig:1}.
Then the equation of state, which determines the magnetization $m$ as a
function of $t$ and the magnetic field $h$, can be written
\begin{equation}
tm + m^{d/2} + m^3 = h\quad,
\label{eq:2.1}
\end{equation}
where we have left out all prefactors of the various terms. 
Equation (\ref{eq:2.1}) is valid for all dimensions $d>2$. Notice the
term $m^{d/2}$, which occurs in addition to what otherwise is an ordinary
mean-field equation of state. It is a manifestation of the soft particle-hole
excitations mentioned in the Introduction. For $d<6$ it dominates the 
$m^3$-term, and
hence we have for the exponent $\beta$, which determines the vanishing of
the zero-field magnetization via $m(t,h=0) \sim t^{\beta}$,
\begin{mathletters}
\label{eqs:2.2}
\begin{equation}
\beta = \cases{2/(d-2)& for $2<d<6$\cr%
                   1/2& for $d>6$\cr}%
     \quad.
\label{eq:2.2a}
\end{equation}
Similarly, the exponent $\delta$, defined by $m(t=0,h) \sim h^{1/\delta}$,
is obtained as
\begin{equation}
\delta = \cases{d/2& for $2<d<6$\cr%
                  3& for $d>6$\cr}%
     \quad.
\label{eq:2.2b}
\end{equation}
\end{mathletters}%

Now let us consider the order parameter field $M({\bf x},t)$ as a function
of space and time, i.e. the field whose average yields the magnetization,
$\langle M({\bf x},t)\rangle = m$. Here the angular brackets 
$\langle\ldots\rangle$ denote a trace with the full statistical operator,
i.e. they include a quantum mechanical expectation value, a disorder
average, and at nonzero temperature also a thermal average. We first
consider the case of $T=0$, and Fourier
transform to wave vectors ${\bf q}$ (with modulus $q=\vert{\bf q}\vert$)
 and frequencies $\omega$. For the order parameter correlation function 
$G(q,\omega) = \langle M({\bf q},\omega)\,M(-{\bf q},-\omega)\rangle$
we find in the limit of small $q$ and $\omega$,
\begin{equation}
G(q,\omega) = {1\over t + q^{d-2} + q^2 - i\omega/q^2}\quad.
\label{eq:2.3}
\end{equation}
Here we have again omitted all prefactors of the terms in the denominator,
since they are of no relevance for our discussion.
The most interesting feature in Eq.\ (\ref{eq:2.3}) is the term $q^{d-2}$.
It is again an immediate consequence of the additional soft modes discussed
in the first section, and Eq.\ (\ref{eq:2.3}), like Eq.\ (\ref{eq:2.1}),
is valid for $d>2$. For $q=\omega=0$, the correlation function $G$ determines
the magnetic susceptibility $\chi_m \sim G(q=0,\omega =0)$ 
in zero magnetic field.
Hence we have $\chi_m (t) \sim t^{-1} \sim t^{-\gamma}$, where the last
relation defines the critical exponent $\gamma$. This yields
\begin{equation}
\gamma = 1\quad,
\label{eq:2.4}
\end{equation}
which is valid for all $d>2$. $\gamma$ thus has its usual mean-field value.
However, for nonzero $q$ the anomalous $q^{d-2}$ term dominates the usual
$q^2$ dependence for all $d<4$. The correlation function at zero frequency
can then be written
\begin{mathletters}
\label{eqs:2.5}
\begin{equation}
G(q,\omega=0) \sim {1\over 1 + (q\xi)^{d-2}}\quad,
\label{eq:2.5a}
\end{equation}
with the correlation length $\xi \sim t^{1/(d-2)} \sim t^{-\nu}$. For
$d>4$ the $q^2$ term is dominant, and we have for the correlation length
exponent $\nu$,
\begin{equation}
\nu=\cases{1/(d-2)& for $2<d<4$\cr%
               1/2& for $d>4$\cr}%
     \quad,
\label{eq:2.5b}
\end{equation}
\end{mathletters}%
Note that $\nu \geq 2/d$, as it must be according to the discussion in the
Introduction. The wavenumber dependence of $G$ at criticality, 
i.e. at $t=0$, is
characterized by the exponent $\eta$: $G(q,\omega=0) \sim q^{-2+\eta}$.
From Eq.\ (\ref{eq:2.3}) we obtain,
\begin{equation}
\eta = \cases{4-d& for $2<d<4$\cr%
                0&   for $d>4$\cr}%
        \quad,
\label{eq:2.6}
\end{equation}
Finally, consider the correlation function at a wavenumber such that
$q\xi = 1$. Then it can be written 
\begin{mathletters}
\label{eqs:2.7}
\begin{equation}
G(q=\xi^{-1},\omega) \sim {1\over 1 - i\omega\tau}\quad,
\label{eq:2.7a}
\end{equation}
with the relaxation or correlation time $\tau \sim \xi^2/t \sim \xi^{2+1/\nu}
\sim \xi^z$, where the last relation defines the dynamical critical exponent
$z$. From Eq.\ (\ref{eq:2.5b}) we thus obtain,
\begin{equation}
z = \cases{d& for $2<d<4$\cr%
           4& for $d>4$\cr}%
        \quad.
\label{eq:2.7b}
\end{equation}
\end{mathletters}%
Notice that with increasing dimensionality $d$, the exponents $\nu$, $\eta$, 
and $z$ `lock into' their mean-field values at $d=d_c^{++}=4$, while $\beta$
and $\delta$ do so only at $d=d_c^{+++}=6$. 
In the special dimensions $d=4$ and $d=6$ the power law scaling behavior
quoted above holds only up to additional multiplicative logarithmic
dependences on the variables $t$, $h$, and $T$. Since these corrections to
scaling occur only in unphysical dimensions they are of academic interest
only, and we refer the interested reader to Refs.\ \onlinecite{fm} for
details.

The results for the clean case are qualitatively similar, but the anomalous
term in the equation of state, Eq.\ (\ref{eq:2.1}), is $m^d$ instead of
$m^{d/2}$. This is because the additional soft modes in that case are
ballistic instead of diffusive, so their frequency scales with wavenumber
like $\omega \sim q$ rather than $\omega \sim q^2$. As a result, the two
special dimensions $d_c^{++}$ and $d_c^{+++}$ coincide,
and are now $d_c^{++}=3$,
while the upper critical dimension proper, above which fluctuations are
irrelevant, is $d_c^+=1$. For $1<d<3$, the exponent values are
$\beta = \nu = 1/(d-2)$, $\delta = z = d$, $\eta = 3-d$, and $\gamma = 1$.
For $d>3$, all exponents take on their mean-field values as they do in the
disordered case for $d>6$, and in $d=3$ there are logarithmic corrections
to power-law scaling.

We now turn to the behavior at nonzero temperatures. Then the equation of
state acquires temperature corrections, and it is helpful to distinguish
between the cases $m>>T$ and $m<<T$, with $m$ and $T$ measured in suitable
units. Taking into account the leading corrections in either limit,
the equation of state reads
\begin{eqnarray}
tm + m^{d/2}\left(1 + T/m\right) = h\qquad ({\rm for}\quad m>>T)\quad,
\nonumber\\
\left(t + T^{(d-2)/2}\right)m + m^3 = h\qquad ({\rm for}\quad T>>m)\quad.
\label{eq:2.8}
\end{eqnarray}
Equation (\ref{eq:2.8}) shows that for any nonzero temperature the asymptotic
critical behavior is not given by the quantum critical exponents. Since
Eq.\ (\ref{eq:2.8}) takes
temperature into account only perturbatively, it correctly
describes only the initial deviation from the quantum critical behavior, and
approximates the classical critical behavior by the mean-field result. 
A full crossover calculation would yield instead the classical Heisenberg 
critical behavior in the asymptotic limit. Also, we are considering only the
saddle point contribution to the magnetization. For models with no additional
soft modes it has been shown that fluctuations that act as dangerous
irrelevant variables introduce another temperature
scale that dominates the one obtained from the saddle 
point.\cite{Millis,Sachdev96} In the present case, however, fluctuations
are suppressed by the long-range nature of the effective field theory,
and the fluctuation temperature scale is subdominant. The behavior described by
Eq.\ (\ref{eq:2.8}) can be summarized by means of a generalized homogeneity
law,
\begin{mathletters}
\label{eqs:2.9}
\begin{equation}
m(t,T,H) = b^{-\beta/\nu}\,m(tb^{1/\nu}, Tb^{\phi/\nu}, 
                                         Hb^{\delta\beta/\nu})\quad.
\label{eq:2.9a}
\end{equation}
Here $\beta$, $\nu$, and $\delta$ have the values given above, and
$b$ is an arbitrary scale factor.
\begin{equation}
\phi = 2\nu\quad,
\label{eq:2.9b}
\end{equation}
\end{mathletters}%
is the crossover exponent that describes the deviation from the quantum
critical behavior due to the relevant perturbation provided by the nonzero
temperature. The entry $Tb^{\phi/\nu} = Tb^2$ in the scaling function in
Eq.\ (\ref{eq:2.9a}) reflects the fact that the temperature dependence of the
saddle point solution is determined by that of the diffusive modes, i.e.
frequency or temperature scales like $T \sim q^2 \sim b^{-2}$. The critical
temperature scale, $T \sim b^{-z}$, would be dominant if it were present, but
since the leading behavior of the magnetization is not determined by
critical fluctuations, it is suppressed.

By differentiating Eq.\ (\ref{eq:2.9a}) with respect to the
magnetic field $h$, one obtains an analogous homogeneity law for the
magnetic susceptibility, $\chi_m$,
\begin{mathletters}
\label{eqs:2.10}
\begin{equation}
\chi_m(t,T,H) = b^{\gamma/\nu}\,\chi_m(tb^{1/\nu}, Tb^{\phi/\nu}, 
                                               Hb^{\delta\beta/\nu})\quad,
\label{eq:2.10a}
\end{equation}
with
\begin{equation}
\gamma = \beta (\delta -1) = 1\quad,
\label{eq:2.10b}
\end{equation}
in agreement with Eq.\ (\ref{eq:2.4}). This result is in agreement with
a more direct calculation of $\chi_m$: The same temperature corrections
that modify the equation of state, Eq.\ (\ref{eq:2.8}), lead to a
replacement of the term $q^{d-2}$ in the denominator of Eq.\ (\ref{eq:2.3})
by $(q^2 + T)^{(d-2)/2}$. Since the homogeneous order parameter correlation
function determines the spin or order parameter susceptibility, this yields
\begin{equation}
\chi_m(t,T) = {1\over t + T^{1/2\nu}}\quad,
\label{eq:2.10c}
\end{equation}
\end{mathletters}%
in agreement with Eqs.\ (\ref{eq:2.10a},\ \ref{eq:2.10b}).

Finally, the critical behavior of the specific heat $c_V$ has been calculated.
It is most convenient to discuss the specific heat coefficient,
$\gamma_V = \lim_{T\rightarrow 0} c_V/T$, which in a Fermi liquid 
would simply be a constant. Its behavior at criticality,
$t=0$, is adequately represented by the integral
\begin{mathletters}
\label{eqs:2.11}
\begin{equation}
\gamma_V = \int_0^{\Lambda} dq\ {q^{d-1}\over T+q^d+q^4+h^{1-1/\delta}q^2}\quad.
\label{eq:2.11a}
\end{equation}
Remarkably, in zero magnetic field, $\gamma_V$ diverges logarithmically as 
$T\rightarrow 0$ for all dimensions $2<d<4$. This can be shown to be a
consequence of the dynamical exponent $z$ being exactly equal to the spatial
dimensionality $d$ in that range of dimensionalities. If one restores the
dependence of $\gamma_V$ on $t$, 
then one obtains a generalized homogeneity law
with a logarithmic correction for the leading scaling behavior of $\gamma_V$,
\begin{eqnarray}
\gamma_V(t,T,H) = \Theta(4-d)\,\ln b \qquad\qquad\qquad\qquad
\nonumber\\
            + F_{\gamma}(t\,b^{1/\nu},T\,b^z,
                  H\,b^{\delta\beta/\nu})\quad.
\label{eq:2.11b}
\end{eqnarray}
\end{mathletters}%
Here $\Theta(x)$ denotes the step function, and $F_{\gamma}$ is an unknown
scaling function. Note that $\gamma_V$ is determined by Gaussian fluctuations
and depends on the critical temperature scale, i.e. $T$ scales like
$t^{\nu z}$ in Eq.\ (\ref{eq:2.11b}). This is the leading temperature
scale, and whenever it is present it dominates the diffusive temperature scale
that shows in Eqs.\ (\ref{eqs:2.9}) and (\ref{eqs:2.10}).

In the clean case, Eqs.\ (\ref{eq:2.9a}) and (\ref{eqs:2.10}) still hold,
if one uses the appropriate exponent values and replaces Eq.\ (\ref{eq:2.9b})
by $\phi=\nu$. In Eq.\ (\ref{eq:2.11a}), the term $q^4$ in the denominator
of the integrand gets replaced by $q^3$, and consequently the argument of
the $\Theta$-function in Eq.\ (\ref{eq:2.11b}) is $3-d$ rather than $4-d$.

\section{Experimental Implications, and Discussion}
\label{sec:III}

\subsection{Experimental Implications}
\label{subsec:III.A}

\begin{figure}
\epsfxsize=8.25cm
\epsfysize=6.6cm
\epsffile{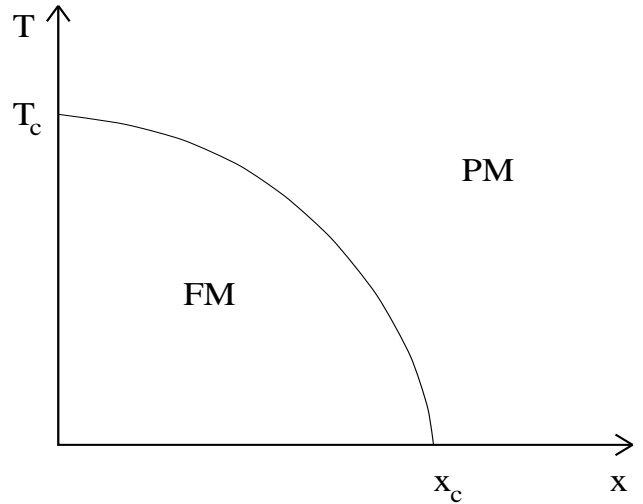}
\vskip 0.5cm
\caption{Schematic phase diagram for an alloy of the form
${\rm P}_x\,{\rm F}_{1-x}$. $T_c$ is the Curie temperature for the pure
ferromagnet F, and $x_c$ is the critical concentration.}
\label{fig:3}
\end{figure}

Let us now discuss the experimental implications of the results presented
in the preceding section. Obviously, one needs a material that shows a
transition from a paramagnetic state to a ferromagnetic one at zero
temperature as a function of some experimentally tunable parameter $x$.
Obvious candidates are magnetic alloys of the stoichiometry 
${\rm P}_x{\rm F}_{1-x}$, with P a paramagnetic metal and F a ferromagnetic 
one. Such materials show the desired transition as a function of the
composition parameter $x$; examples include Ni for the ferromagnetic
component, and Al or Ga for the paramagnetic one.\cite{Mott} At the critical
concentration $x_c$ they also are substantially disordered, but due to the fact
that both constituents are metals they are far from any metal-insulator
transition. Our theory should therefore be applicable to these systems. 
The schematic phase diagram at $T=0$ in the $T$-$x$ plane
is shown in Fig.\ \ref{fig:3}. Notice that this is a realistic
phase diagram, as opposed to
the `theoretical' ones in Figs.\ \ref{fig:1} and \ref{fig:2}. 
A change of the composition parameter $x$ leads, besides a change of $F_0^a$, 
to many other changes in the microscopic parameters of the system. As $x$ is
varied, the system will therefore move on a complicated path in the diagram
shown in, say, Fig.\ \ref{fig:1}. However, 
since the critical behavior near the transition is universal, it is independent
of the exact path traveled.

\begin{figure}
\epsfxsize=8.25cm
\epsfysize=6.6cm
\epsffile{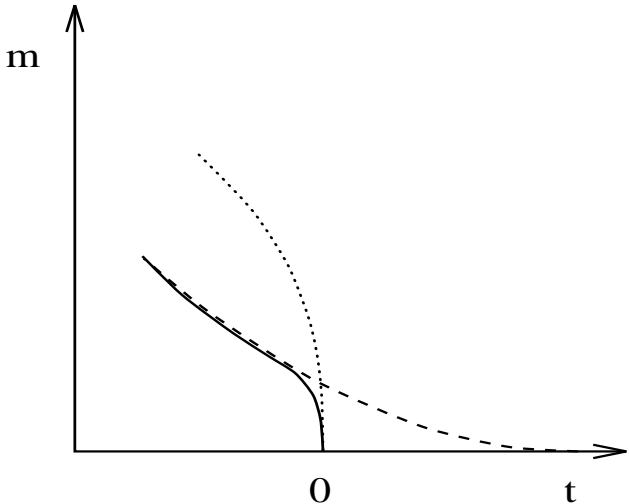}
\vskip 0.5cm
\caption{Schematic critical behavior of the magnetization $m$ at nonzero
temperature, showing the crossover from the quantum critical behavior
($\beta=2$, dashed line) to the classical critical behavior 
($\beta\approx 0.37$, dotted line).
Notice that the actual transition is classical in nature.}
\label{fig:4}
\end{figure}

One possible experiment would consist in driving the system at a low, fixed
temperature through the transition by changing the composition $x$.
While this involves the preparation of many samples, this way of probing a
quantum phase transition has been used to observe the metal-insulator
transition in P doped Si.\cite{vL} It might also be possible to use the
stress tuning technique that has been used for the same purpose.\cite{stress}
Either way one will cross the transition line along a more or less vertical
path in Fig.\ \ref{fig:2}, and for a sufficiently low temperature this path
will go through both the classical and the quantum critical region indicated
in the inset in Fig.\ \ref{fig:2}. Due to the large difference between the
quantum critical exponents quoted in Sec.\ \ref{sec:II} and the corresponding
exponents for classical Heisenberg magnets, the resulting crossover should
be very pronounced and easily observable. For instance, for $3$-$d$ systems
our Eq.\ (\ref{eq:2.2a}) predicts $\beta = 2$, while the value for the
thermal transition is $\beta_{\rm class}\approx 0.37$. The resulting
crossover in the critical behavior of the magnetization is schematically
shown in Fig.\ \ref{fig:4}. Alternatively, 
one could prepare a sample with a
value of $x$ that is as close as possible to $x_c$, and measure the
magnetic field dependence of the magnetization, extrapolated to $T=0$, to
obtain the exponent $\delta$. Again, there is a large difference between
our prediction of $\delta=1.5$ in $d=3$, and the classical value
$\delta_{\rm class}\approx 4.86$.

\begin{figure}
\epsfxsize=8.25cm
\epsfysize=6.6cm
\epsffile{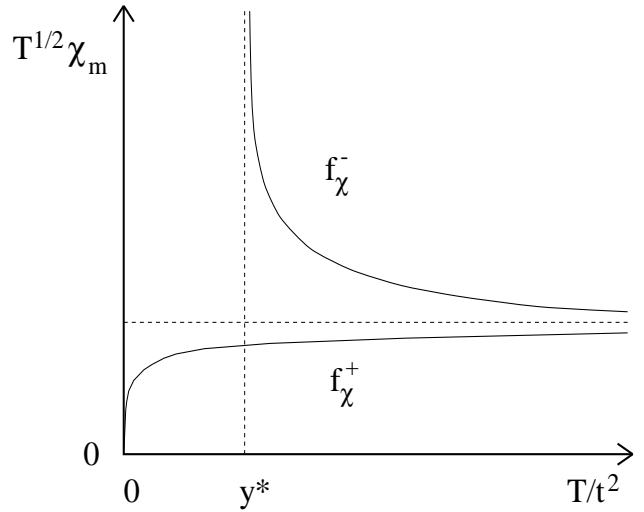}
\vskip 0.5cm
\caption{Schematic prediction for a scaling plot of the magnetic
suscteptibility.}
\label{fig:5}
\end{figure}

Yet another possibility
is to measure the zero-field magnetic susceptibility as a function of both
$t=\vert x-x_c\vert$ and $T$. Equation (\ref{eq:2.10a}) predicts
\begin{equation}
\chi_m(t,T) = T^{-1/2}\,f_{\chi}(T/t^2)\quad.
\label{eq:3.1}
\end{equation}
Here $f_{\chi}$ is a scaling function that has two branches, $f_{\chi}^+$
for $x>x_c$, and $f_{\chi}^-$ for $x<x_c$. Both branches approach a constant
for large values of their argument,
$f_{\chi}^{\pm}(y\rightarrow\infty)={\rm const.}$ For small arguments, we
have $f_{\chi}^+(y\rightarrow 0)\sim \sqrt{y}$, while $f_{\chi}^-$ diverges
at a nonzero value $y^*$ of its argument that signalizes the classical
transition, $f_{\chi}^-(y\rightarrow y^*)\sim (y-y^*)^{-\gamma_{\rm class}}$,
with $\gamma_{\rm class}\approx 1.39$ the susceptibility exponent for
the classical transition. Our prediction is then that a plot of
$\chi_m\ T^{1/2}$ versus $T/t^2$ will yield a universal function the
shape of which is schematically shown in Fig.\ \ref{fig:5}.
Notice that the exponents are known {\it exactly}, so the only adjustable
parameter for plotting experimental data will be the position of the
critical point. This is on sharp contrast to some other quantum phase
transitions, especially metal-insulator transitions, where the exponent
values are not even approximately known, which makes scaling plots almost
meaningless.\cite{glass}
 
Finally, one can consider the low-temperature behavior of the specific
heat. According to Eq.\ (\ref{eq:2.11b}), as the temperature is lowered
for $x\agt x_c$ the leading temperature dependence of
the specific heat will be
\begin{mathletters}
\label{eqs:3.2}
\begin{equation}
c_V(T) \sim T\ln T\quad.
\label{eq:3.2a}
\end{equation}
At criticality this behavior will continue to $T=0$, while for $x>x_c$
it will cross over to
\begin{equation}
c_V(T) \sim (\ln t)\ T\quad.
\label{eq:3.2b}
\end{equation}
\end{mathletters}%
For $x\alt x_c$ one will encounter the classical Heisenberg transition
where the specific heat shows a finite cusp (i.e., the exponent $\alpha$,
defined by $c_V \sim (T-T_c)^{-\alpha}$, is negative).

\subsection{Theoretical Discussion}
\label{subsec:III.B}

There are also various theoretical implications of the results presented
in Sec.\ \ref{sec:II}. One aspect is the general message that the usual
LGW philosophy must not be applied uncritically to quantum phase transitions,
because of the large number of soft modes that exist at zero temperature
in a generic system. If any of these couple to the order parameter, then
an effective theory entirely in terms of the order parameter will not be
well behaved. In the present case we have actually been able to use this
to our advantage, since the long-ranged interaction that the additional
soft modes induce in the order parameter theory suppress the disorder
fluctuations, which is the reason for the remarkable fact that we are
able to exactly determine the critical behavior of a three-dimensional,
disordered system. In general, however, the presence of soft modes in
addition to the order parameter fluctuations will call for the derivation
of a more complete low-energy effective theory that keeps {\em all} of the
soft modes explicitly.

\begin{figure}
\epsfxsize=8.25cm
\epsfysize=6.6cm
\epsffile{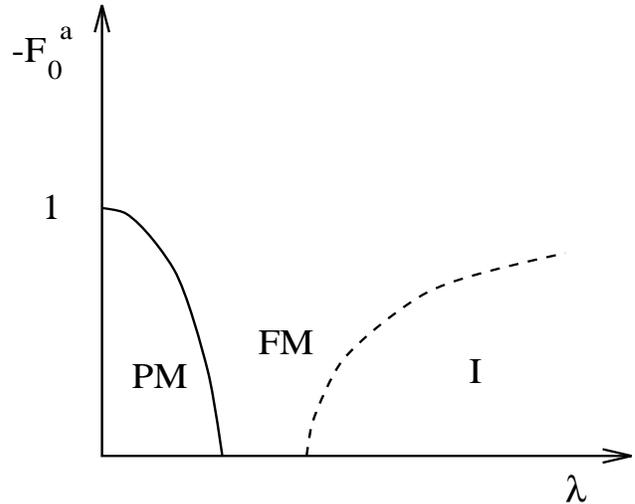}
\vskip 0.5cm
\caption{Schematic phase diagram for a disordered itinerant electron
system at $T=0$ close to $d=2$. The phases shown are the paramagnetic
metal (PM), the ferromagnetic metal (FM), and the insulator (I) phase.
Whether within I there is another phase transition from a ferromagnetic
to a paramagnetic insulator is not known.}
\label{fig:6}
\end{figure}

Another very interesting aspect is a connection between our results on
the ferromagnetic transition, and a substantial body of literature on a
problem that appears in the theory of the metal-insulator 
transition in interacting disordered electron systems, i.e. the transition
from PM to PI in Fig.\ \ref{fig:1}. This problem has been known ever since
the metal-insulator transition of interacting disordered electrons was
first considered, and it has led to substantial confusion in that field.
Early work on the metal-insulator transition
showed that in two-dimensional systems without impurity
spin-flip scattering, the spin-triplet interaction amplitude
scaled to large values
under renormalization group iterations.\cite{RunawayFlow} This is still
true in $d=2+\epsilon$, and since the runaway flow occurs before the
metal-insulator transition is reached, this precluded the theoretical
description of the latter in such systems. This problem was
interpreted, incorrectly as it turned out later, as a signature of local
moment formation in all dimensions.\cite{LM}
Subsequently, the present authors
studied this problem in some detail.\cite{IFS} We were able to explicitly
resum the perturbation theory and show that at a critical value of the
interaction strength, or of the disorder, there is a bulk, thermodynamic
phase transition in $d>2$ that is {\em not} the metal-insulator transition. 
While this ruled out local moments (which would not lead to to phase
transition), the physical meaning of this transition was obscure
at the time since no order parameter had been identified, and its description
was entirely in terms of soft diffusion modes. However, the critical exponents
obtained are identical to those given in Sec.\ \ref{sec:II} 
for the quantum ferromagnetic
phase transition, and in both cases logarithmic corrections to scaling
are found.\cite{logfootnote}
Because the exponents in the two cases are identical, we
conclude that the transition found earlier by us, whose physical nature
was unclear, is actually the ferromagnetic transition.
One also concludes that our speculations in Ref.\ \onlinecite{IFS}
about the nature of the ordered phase as an
`incompletely frozen spin phase' with no long-range magnetic order, were
not correct; this phase is actually the metallic ferromagnetic 
phase. On the other hand, the techniques used in that reference
allowed for a determination of the qualitative phase diagram as a function
of dimensionality, which our present analysis is not capable of. 
This analysis showed the existence of yet another interesting dimensionality
above $d=2$, which we denote by $d^*$. With
the appropriate reinterpretation of the `incompletely frozen spin phase'
as the ferromagnetic phase, the qualitative phase diagram for $2<d<d^*$ is
shown in Fig.\ \ref{fig:6}. Compared to Fig.\ \ref{fig:1}, the following
happens as $d$ is lowered from $d=3$: The multicritical point M moves to
downward, and at $d=d^*$ it reaches the $\lambda$-axis. $d^*$ was
estimated in Ref.\ \onlinecite{IFS} to be approximately $d^*=2.03$. For $d<d^*$,
the insulator phase can therefore not be reached directly from the
paramagnetic metal. This explains why in the perturbative renormalization
group calculations in $d=2+\epsilon$ one necessarily encounters the
ferromagnetic transition first, and it should finally put to rest the long
discussion about the physical meaning of the runaway flow that is encountered
in these theories. It also shows that
none of these theories are suitable
for studying the metal-insulator transition in the absence of spin-flip
mechanisms, as they start out in the wrong phase.

It should also be pointed out that our earlier theory depended crucially
on there being electronic spin conservation. This feature would be lost
of there were some type of impurity  spin flip scattering process. In
that case, the soft modes that lead to the long-range order parameter
interactions acquire a mass or energy gap, and at sufficiently large
scales the interactions are effectively of short range. The asymptotic
critical phenomena in this case are described by a short-range, local
order parameter field theory with a random mass, or temperature, term.
Such a term is present in the case of a conserved order parameter also,
but due to the long ranged interaction it turns out to be irrelevant 
with respect to the nontrivial
Gaussian fixed point. In the absence of the conservation law, however,
the random mass term is relevant with respect to
the Gaussian fixed point analogous to the one discussed here. This
underscores the important role that is played by the order parameter
being conserved in our model. The quantum phase transition in a model
where it is not has been discussed in Ref.\ \onlinecite{afm}.

We finally discuss why some of our results are in disagreement with
Sachdev's\cite{Sachdev} general scaling analysis of quantum phase
transitions with conserved order parameters. For instance, it follows from
our Eqs.\ (\ref{eqs:2.10},\ \ref{eq:2.11b}) that the Wilson ratio, defined as
$W = (m/H)/(C_V/T)$, diverges at criticality rather than being a universal
number as predicted in Ref.\ \onlinecite{Sachdev}. Also, for $2<d<4$ the
function $F_{\gamma}$ in Eq.\ (\ref{eq:2.11b}), for $t=0$ and neglecting
corrections to scaling, is a function of $T/H$, in agreement with
Ref.\ \onlinecite{Sachdev}, but for $d>4$ this is not the case.
The reason for this breakdown of general scaling is that we work
above an upper critical dimensionality, and hence dangerous irrelevant
variables\cite{MEF} 
appear that prevent a straightforward application of the results 
of Ref.\ \onlinecite{Sachdev} to the present problem. These dangerous
irrelevant variables have to be considered very carefully,
and on a case by case
basis. This caveat is particularly relevant for quantum phase transitions
since they tend to have a low upper critical dimension.
It is well known that a given irrelevant variable can be
dangerous with respect to some observables but not with respect to others.
Specifically, in our case there is a dangerous irrelevant variable
that affects the leading scaling behavior of the magnetization,
but not that of the specific heat coefficient, and this leads to the divergence
of the Wilson ratio. This dangerous irrelevant variable is also the reason 
why the exponents $\beta$ and
$\delta$, which describe the critical behavior of the magnetization, remain
dimensionality dependent up to $d=6$, while all other exponents `lock into'
their mean-field values already at $d=4$.

\section{Theoretical Outline}
\label{sec:IV}

Here we sketch the derivation of the results that were presented in Sec.\ 
\ref{sec:II}. We do so for completeness only, and will be very brief. A
detailed account of the derivation can be found in Ref.\ \onlinecite{fm}
for the disordered case, and in Ref.\ \onlinecite{clean} for the clean case.

Hertz\cite{Hertz} has shown how to derive an LGW functional for a quantum
ferromagnet. One starts by separating the spin-triplet part
of the electron-electron interaction, i.e. the interaction between spin
density fluctuations, from the rest of the action, writing
\begin{mathletters}
\label{eqs:4.1}
\begin{equation}
S = S_0 + S_{int}^{\,(t)}\quad,
\label{eq:4.1a}
\end{equation}
with
\begin{equation}
S_{int}^{\,(t)} = {\Gamma_t\over 2} \int dx\,{\bf n}_s(x)
                 \cdot {\bf n}_s(x)\quad.
\label{eq:4.1b}
\end{equation}
\end{mathletters}%
Here $S_{int}^{\,(t)}$ is the spin-triplet interaction part of the action,
and $S_0$ contains all other parts, in particular the electron-electron
interaction in all other channels. $\Gamma_t$ is the spin triplet interaction
amplitude, which is related to the Landau paramter $F_0^a$ used above by
$\Gamma_t = -F_0^a/(1 + F_0^a)$, ${\bf n}_s(x)$ is the electron spin density
vector, $x = ({\bf x},\tau)$ denotes space and imaginary time, and
$\int dx = \int d{\bf x}\int_0^{1/T} d\tau$. In the critical region near a
quantum phase transition, imaginary time scale like a length to the power
$z$, and the space-time nature of the integrals in the action accounts for
the system's effective dimension $d+z$ that was mentioned in the Introduction.

Now $S_{int}^{\,(t)}$ is
decoupled by means of a Hubbard-Stratonovich transformation.\cite{Hertz}
The partition function, apart from a noncritical multiplicative constant,
can then be written
\begin{mathletters}
\label{eqs:4.2}
\begin{equation}
Z = \int D[{\bf M}]\ \exp\left(-\Phi[{\bf M}]\right)\quad,
\label{eq:4.2a}
\end{equation}
with the LGW functional
\begin{eqnarray}
\Phi[&&{\bf M}] = {\Gamma_t\over 2} \int dx\ {\bf M}(x)\cdot {\bf M}(x)
                \quad\quad\quad\quad\quad\quad
\nonumber\\
              &&- \ln \left<\exp\left[-\Gamma_t \int dx\ {\bf M}(x)\cdot
                              {\bf n}_s(x)\right]\right>_{S_0}\quad.
\label{eq:4.2b}
\end{eqnarray}
\end{mathletters}%
Here $\langle\ldots\rangle_{S_0}$ denotes an average taken with the action
$S_0$. If the LGW functional $\Phi$ is formally expanded in powers of 
${\bf M}$, then the term of order ${\bf M}^n$ obviously has a coefficient 
that is given by a connected $n$-point spin density correlation function
of the `reference system' defined by the action $S_0$. 

At this point we need to remember that our reference system $S_0$ contains
quenched disorder, which has not been averaged over yet.
The $n$-point correlation functions that form the coefficients of the
LGW functional therefore still
depend explicitly on the particular realization of the randomness in the
system. The average over the quenched disorder, which we denote by
$\{\ldots \}_{\rm dis}$, requires averaging the
free energy, i.e. we are interested in $\{\ln Z\}_{\rm dis}$.
This is most easily done by means of the replica trick,\cite{Grinstein},
i.e. one writes 
\begin{eqnarray}
\{\ln Z\}_{\rm dis} = \lim_{n\rightarrow 0}\,{1\over n}\,
                                   \left[\{Z^n\}_{\rm dis} - 1\right]
\nonumber\\
 = \lim_{n\rightarrow 0}\,{1\over n}\,
                        \left[\int\prod_{\alpha} D[{\bf M}^{\alpha}]\left\{
    e^{-\sum_{\alpha = 1}^n \Phi^{\alpha}[{\bf M}^{\alpha}]}\right\}_{\rm dis}
    - 1\right]\ ,
\label{eq:4.3}
\end{eqnarray}
where the index $\alpha$ labels $n$ identical replicas of the system.
The disorder average is now easily performed by expanding the exponential
in Eq.\ (\ref{eq:4.3}). Upon reexponentiation, the coefficients in the
replicated LGW functional are disorder averaged correlation functions 
of the reference system that are cumulants with respect to the disorder
average. The Gaussian part of $\Phi^{\alpha}$ is simply
\begin{eqnarray}
\Phi_{(2)}^{\alpha}[{\bf M}^{\alpha}] = {1\over 2} \int dx_1\,dx_2\,
              {\bf M}^{\alpha}(x_1)\biggl[\delta(x_1 - x_2)
\nonumber\\
- \Gamma_t \chi(x_1 - x_2)\biggr]\cdot {\bf M}^{\alpha}(x_2)\quad.
\label{eq:4.4}
\end{eqnarray}
Here $\chi(x)$ is the disorder averaged spin susceptibility or 2-point
spin density correlation function of the
reference system. The cubic term, $\Phi_{(3)}^{\alpha}$ has a coefficient
given by the averaged 3-point spin density correlation function. For the
quartic term, the cumulant nature of these correlation functions leads
to two terms with different replica structures, and higher order terms
have correspondingly more complicated structures.

The next step is to calculate the spin density correlation functions for
the reference system. It now becomes important that we have
kept in our action $S_0$ the electron-electron interaction in all channels
except for the spin-triplet one that has been decoupled in deriving the
LGW functional. At this point our treatment deviates from that of Hertz,
who took the reference ensemble to describe noninteracting electrons. This was
generally considered an innocent approximation that should not have any
qualitative effects. However, this belief was mistaken, since the spin 
density correlations of interacting
electrons are qualitatively different from those of noninteracting ones.
The spin susceptibility can be easily calculated in perturbation theory.
The result shows that the static spin susceptibility as a function of
the wavenumber $q$ is nonanalytic at $q=0$. For small $q$ it has the
form 
\begin{equation}
\chi(q) = {\rm const} - q^{d-2} - q^2 \quad.
\label{eq:4.5}
\end{equation}
The nonanalyticity is a
consequence of the presence of soft particle-hole excitations in the
spin-triplet channel, and it occurs only in an interacting electron
system. That is, the prefactor of the $q^{d-2}$ term, which we have
suppressed in Eq.\ (\ref{eq:4.5}), 
vanishes for vanishing interaction amplitudes. 
Renormalization group arguments can then be used to ascertain that
this perturbative result indeed represents the exact behavior of $\chi$ in
the long-wavelength limit. 
If one also considers the frequency dependence of $\chi$, one
obtains the Gaussian part of the LGW functional in the form
\begin{mathletters}
\label{eqs:4.6}
\begin{eqnarray}
\Phi^{\alpha}_{(2)}[{\bf M}] &&= {1\over 2} \sum_{\bf q}\sum_{\omega_n}
     {\bf M}^{\alpha}({\bf q},\omega_n)\,\left[t_0 + q^{d-2} \right.
\nonumber\\
      &&\left. +q^2 + \vert\omega_n\vert/q^2\right]\,
     \cdot {\bf M}^{\alpha}(-{\bf q},-\omega_n)\ ,
\label{eq:4.6a}
\end{eqnarray}
where
\begin{equation}
t_0 = 1 - \Gamma_t\,\chi_s({\bf q}\rightarrow 0,\omega_n=0)\quad,
\label{eq:4.6b}
\end{equation}
\end{mathletters}%
is the bare distance from the critical point, and the $\omega_n = 2\pi Tn$
are bosonic Matsubara frequencies.

The Gaussian theory, Eqs.\ (\ref{eqs:4.6}), can be analyzed using standard
renormalization group techniques.\cite{Ma} Such an analysis reveals
the existence of a Gaussian fixed point whose critical properties are
the ones given in Sec.\ \ref{sec:II}. The remanining question is whether
this fixed point is stable with respect to the higher, non-Gaussian terms
in the action. These higher terms also need to be considered in order to
obtain the critical behavior of the magnetization.

A calculation of the higher correlation functions that determine the
non-Gaussian vertices of the field theory shows that the nonanalyticity
that is analogous to the one in the spin susceptibility, Eq.\ (\ref{eq:4.5}),
is stronger and results in a divergence of these correlation functions in
the zero frequency, long-wavelength limit. Specifically, the leading
behavior of
the $n$-point spin density correlation that determines the coefficient of
the term of order ${\bf M}^n$ in the LGW functional, considered at vanishing
frequencies as a function of a representative wavenumber $q$, is
\begin{equation}
\chi^{(n)}(q\rightarrow 0) \sim q^{d+2-2n}\quad.
\label{eq:4.7}
\end{equation}
As a result, the coefficients cannot, as usual, be expanded about zero
wavenumber, and the theory is nonlocal. Despite this unpleasant behavior
of the field theory, it is easy to see by power counting that all of
these terms except for one are irrelevant with respect to the Gaussian 
fixed point in all dimensions $d>2$. The one exception is the quartic
cumulant contribution that is the disorder average of the square of the
spin susceptibility, which is marginal in $d=4$, but irrelevant in all
other dimensions. This term is physically of interest, since it represents
the random mass or random temperature contribution that one would expect
in a theory of disordered magnets, and that was mentioned in
Sec.\ \ref{subsec:III.B} above. 

The conclusion from these considerations is that apart from logarithmic
corrections to scaling in certain special dimensions, 
the Gaussian theory yields
the exact critical behavior, and the only remaining question pertains to
the form of the equation of state. Since the quartic coefficient
$\chi^{(4)}$ is a dangerous irrelevant variable for the magnetization,
this requires a scaling interpretation of the infrared divergence of
$\chi^{(4)}$. In Ref.\ \onlinecite{fm} it has been shown that for scaling
purposes the wavenumber $q$ in Eq.\ (\ref{eq:4.7}) can be identified
with the magnetization $m^{1/2}$. This is physically plausible, since
the divergence stems from an integration over soft modes that are
rendered massive by an external magnetic field. Since a nonzero
magnetization acts physically like a magnetic field, it cuts off the
singularity in Eq.\ (\ref{eq:4.7}). With this interpretation of the
singular coefficients, the term of order $m^n$ in the saddle point solution
of the LGW theory has the structure $m^{n-1}\,(m^{1/2})^{d+2-2n} = m^{d/2}$,
which leads to the equation of state as given in Eq.\ (\ref{eq:2.1}).
One might wonder why the magnetic fluctuations in the paramagnetic
phase do not also cut off the singularity in Eq.\ (\ref{eq:4.7}), 
and thus weaken
or even destroy the effects discussed above. While such a cutoff mechanism
does exist, it enters the theory only via the fluctuations, which are
RG irrelevant with respect to the Gaussian fixed point. It therefore shows
only in the corrections to scaling, but not in the leading critical
behavior.

Again, all of these arguments can be repeated for the case without disorder.
The only changes one encounters pertain to the values of various exponents
due to the different character of the soft modes. This leads to the results
quoted in Sec.\ \ref{sec:II}.

\acknowledgments

This work was supported by the NSF under grant numbers 
DMR-96-32978 and DMR-95-10185. Part of the work reviewed here was performed
during the 1995 Workshop on Quantum Phase Transitions at the TSRC in
Telluride, CO. We thank the participants of that workshop for stimulating
discussions. We also would like to acknowledge our collaborators on the
clean case, Thomas Vojta and Rajesh Narayanan.


\begin{references}
\bibitem{WeakLoc} The term `weak localization' is ill-defined, and often
 misunderstood. Here we use it to refer to the
 nonanalytic behavior of electronic correlation functions in the limit
 of zero momentum and/or frequency that is induced by quenched disorder,
 or by a combination of interactions and quenched disorder, and that
 occurs even if the system is far from any kind of metal-insulator
 transition. The physical reason for these nonanalyticities is the diffusive
 motion of the electrons in the presence of quenched disorder; for a
 detailed discussion see, e.g., Refs.\ \onlinecite{LeeRama} and \onlinecite{R}.
\bibitem{logfootnote} The logarithmic corrections to scaling in the two
 theories are not quite the same. We believe this is due to the
 fact that of the two integral equations in Ref.\ \onlinecite{IFS} 
 only one was exact.
\end{references}
\end{document}